\documentclass{elsart}
\usepackage{graphics}
\usepackage{txfonts}
\usepackage{amssymb}
\newcommand{\bc}{\begin{center}}
\newcommand{\ec}{\end{center}}
\newcommand{\be}{\begin{equation}}
\newcommand{\ee}{\end{equation}}
\newcommand{\ba}{\begin{array}}
\newcommand{\ea}{\end{array}}
\newcommand{\bea}{\begin{eqnarray}}
\newcommand{\eea}{\end{eqnarray}}
\def\la{\mathrel{\mathchoice {\vcenter{\offinterlineskip\halign{\hfil
$\displaystyle##$\hfil\cr<\cr\sim\cr}}}
{\vcenter{\offinterlineskip\halign{\hfil$\textstyle##$\hfil\cr
<\cr\sim\cr}}}
{\vcenter{\offinterlineskip\halign{\hfil$\scriptstyle##$\hfil\cr
<\cr\sim\cr}}}
{\vcenter{\offinterlineskip\halign{\hfil$\scriptscriptstyle##$\hfil\cr
<\cr\sim\cr}}}}}
\def\l{\lambda}
\def\mkm{{\mu}\rm{m}}

\def\degr{\hbox{$^\circ$}}
\begin{document}
\runauthor{Voshchinnikov and Das}
\begin{frontmatter}
\title{Modelling interstellar extinction and polarization
with spheroidal grains}
\author[SPB1]{N.V.~Voshchinnikov\thanksref{nvv}}
\author[Pune]{H.K.~Das}

\address[SPB1]{Sobolev Astronomical Institute, St.~Petersburg University,
               St.~Petersburg, 198504 Russia}
\address[Pune]{IUCAA,Post Bag 4, Ganeshkhind, Pune 411 007, India}

\bc
({\small\it Received .. October 2007})
\ec
\thanks[nvv]{To whom all correspondence should be addressed.
Phone: (+7) 812/428 42 63;
Fax: (+7) 812/428 71 29; e-mail:nvv@astro.spbu.ru}
\begin{abstract}

We calculate the wavelength dependence of the
       ratio of the linear polarization degree to extinction
       (polarizing efficiency) $P(\lambda)/A(\lambda)$
from the ultraviolet to near-infrared.
The prolate and oblate particles with aspect ratios from
$a/b=1.1$ up to 10
are assumed to be rotating and partially
aligned with the  mechanism of paramagnetic relaxation
(Davis--Greenstein).
       Size/shape/orientation effects are analyzed.
It is found that the wavelength dependence of $P(\lambda)/A(\lambda)$
is mainly determined by the particle composition and size whereas
the values of $P(\lambda)/A(\lambda)$ depend on the
particle shape, degree and direction of alignment.
\end{abstract}
\end{frontmatter}

\section{Introduction}

Modelling of the
wavelength dependencies of interstellar extinction $A(\l)$
and  linear polarization $P(\l)$ allows one
to obtain  information about such properties
of interstellar grains as size,  composition, shape, etc.
Interstellar polarization also tells us about the structure
of magnetic fields because it arises due to dichroic extinction
of non-spherical grains aligned in large-scale Galactic magnetic
fields \cite{gre68}, \cite{ha04}.
A correlation between observed interstellar extinction and polarization
shows that the same particles are responsible for both phenomena.

Very often interstellar  extinction and  polarization
are modelled separately (e.g., \cite{weid01},  \cite{km95}).
The modelling of these phenomena usually includes consideration of
normalized extinction $E(\lambda-{\rm V})/E\rm (B-V)$ and normalized
polarization given by Serkowski's curve
($P_{\rm max}$ is the maximum degree of polarization  and
$\lambda_{\rm max}$ the wavelength corresponding to it,
$K$ is the coefficient)
$P(\lambda)/P_{\rm max} = \exp [-K \ln^2 (\lambda_{\rm max}/\lambda)]$
(see, e.g., \cite{wcm93} and discussion in \cite{w03}, \cite{nvv2}).
In these cases, the important observational data
(absolute values of extinction and polarization) are ignored and
normalized curves can be fitted using
particles of different composition and slightly different
sizes (see, e.g., discussion in \cite{gre68}, \cite{nvv2}).
For example, the consideration of absolute extinction
gives the possibility to use the cosmic abundances to
constrain grain models (see \cite{vihd06}).
In the case of interstellar polarization it is better
to interpret the wavelength dependence of
polarizing efficiency per unit extinction $P(\lambda)/A(\lambda)$
(ratio of the linear polarization
degree to extinction). This allows one to estimate
the degree and direction of grain alignment and particle shape.
Note that  previous consideration of polarizing efficiency
was restricted by one wavelength band only
(like $P({\rm V})/A({\rm V})$ or $P({\rm K})/A({\rm K})$).

Here, we analyze the extinction and polarization produced by
homogeneous spheroidal particles with different aspect ratios $a/b$
($a$ and $b$ are major and minor axes, respectively).
Spheroidal particles have the important applications in various
fields of science (spheroidal antennas, bacteria and microweeds,
raining drops, etc.). By changing $a/b$ the particles with a shape
varying from close to as sphere ($a/b \simeq 1$) up to
needles (prolate spheroids) or disks (oblate spheroids)
can be studied.
The particles are assumed to be rotating and partially
aligned with the Davis--Greenstein mechanism
(mechanism of paramagnetic relaxation) \cite{dg51}.
We focus on the behaviour of polarizing efficiencies
and consider the normalized extinction $A(\lambda)/A_{\rm V}$
(i.e., ignore the cosmic abundances which will be taken into
account in next paper).
The theory is compared with observations of stars
seen apparently through one interstellar cloud.


\section{Polarizing efficiency: observations}\label{obs}

\begin{figure}[htb]
\centerline{
\resizebox{\hsize}{!}{\includegraphics{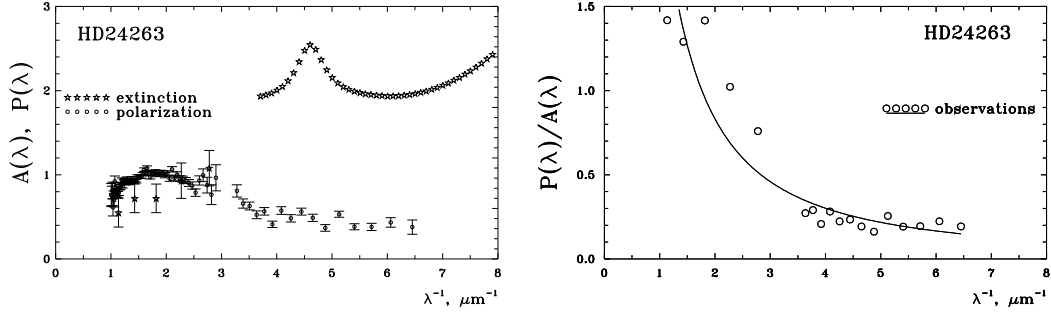}}
}
\caption
{Interstellar extinction (stars and thin line)
and polarization curves for star
HD~24263 (left panel)
and the polarizing  efficiency of the interstellar medium
in the direction of this star (right panel;
solid curve shows the power-law approximation
$P/A \propto \lambda^{1.47}$).
Observational data were taken from \cite{val04} (extinction) and
\cite{andal96} (polarization).
}
\label{fig1}
\end{figure}

Interstellar extinction grows with a radiation wavelength
decrease while interstellar linear polarization has a maximum
in the visual part of spectrum and declines at shorter and
longer wavelengths
(see \cite{w03}, \cite{nvv2} and Fig.~\ref{fig1}, left panel).

The  {\it polarizing efficiency} of the diffuse interstellar medium
is defined as the ratio of the  percentage polarization ($P$)
to the extinction ($A$) observed at the same wavelength
$P(\lambda)/A(\lambda)$.
There exists an empirically found upper limit on this ratio
\begin{equation}
P_{\rm max}/A({\rm V}) \la 3 \%/{\rm mag}\,,
\label{pebv}
\end{equation}
where $P_{\rm max}$ is the maximum degree of linear polarization
which is reached on average near the wavelength
$\lambda_{\rm max} \approx 0.55\,\mu$m
(see \cite{w03}, \cite{nvv2} for more discussion).

\begin{figure}[htb]
\centerline{
\resizebox{12cm}{!}{\includegraphics{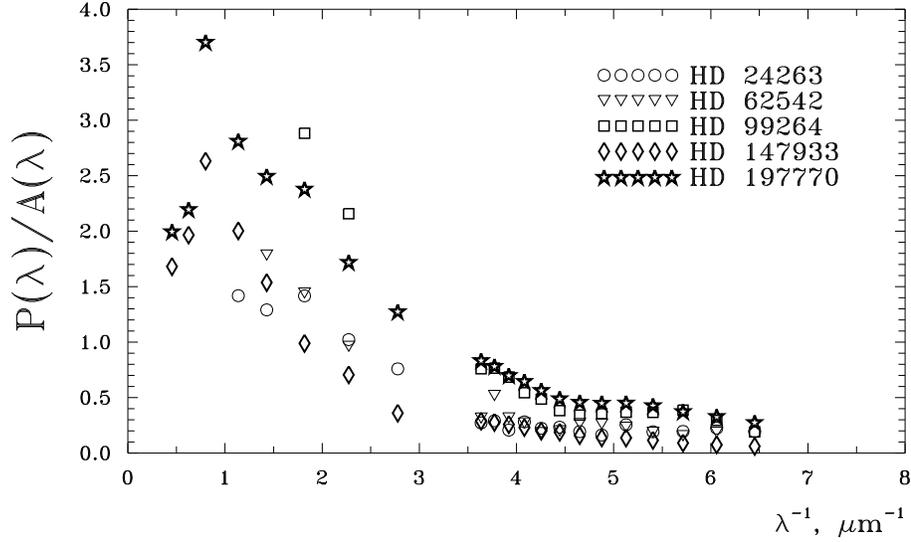}}}
\caption
{Polarizing  efficiency of the interstellar medium
in the direction of five stars.
Observational data were taken from \cite{val04} (extinction) and
\cite{andal96} (UV and visual polarization),
\cite{wilk80} (IR polarization).
}
\label{fig2}
\end{figure}

We chose five stars
located not far than 500~pc
with measured ultraviolet (UV) polarization
\cite{andal96}, then found the extinction (data published
in \cite{val04} were mainly used) and calculated
the ratio $P(\lambda)/A(\lambda)$.
It is suggested that these stars are seen through one interstellar cloud
that is supported by a weak rotation of positional angle \cite{andal96}.
The obtained polarizing efficiencies are shown in
Fig.~\ref{fig2}. Apparently,
first presentation of the observational data in the similar form
was made by Whittet (\cite{wh1}, Fig.~9)  who gave
the average normalized dependence
$P(\lambda)/A(\lambda)\cdot A_{\rm V}/P_{\rm max}$.

Note that the polarizing  efficiency generally increases with a
wavelength growth if $\lambda \la 1\,\mu$m. It may be approximated using
the power-law dependence $P/A \propto \lambda^{\epsilon}$.
For stars presented in Fig.~\ref{fig2} the values of
$\epsilon$ vary from 1.41 for HD~197770 to 2.06 for HD~99264.

\section{Modelling}\label{mod}

Let us consider non-polarized light passing through a cloud of rotating
sphe\-roi\-dal grains.
Interstellar grains are usually partially aligned
(see, e.g., \cite{laz06}).
Imperfect alignment is also supported by the
fact that values of the polarizing   efficiencies calculated
for non-rotating  or  perfectly aligned particles are generally
higher than the empirically estimated upper limit
given by Eq.~(\ref{pebv}) \cite{nvv2}, \cite{nvv1}.

The extinction in stellar magnitudes and
polarization in percentage can be found as \cite{nvv2}
\begin{equation}
A({\lambda})= 1.086 N_{\rm d} \langle C_{\rm ext} \rangle _{\lambda}\,,
\,\,\,\,\,\,\,\,\,\,\,\,\,\,\,
P({\lambda})=N_{\rm d} \langle C_{\rm pol} \rangle_{\lambda}100{\%}\,,
\label{eq2}
\end{equation}
where $N_{\rm d}$ is the dust grain column density and
$ \langle C_{\rm ext} \rangle_{\lambda}$ and
$ \langle C_{\rm pol} \rangle_{\lambda}$ are the extinction and
linear polarization cross sections, respectively, averaged over
the grain orientations
\begin{equation}
 \langle C_{\rm ext} \rangle_{\lambda}={ \left(\frac{2}{\pi} \right)^2}
{\int_{0}^{\pi/2}}{\int_{0}^{\pi/2}}{\int_{0}^{\pi/2}} \pi r_V^2 \,Q_{\rm ext} \,
f(\xi, \beta) \, d{\varphi} d{\omega} d{\beta}\,,
\label{eq3}
\end{equation}
\begin{equation}
 \langle C_{\rm pol} \rangle _{\lambda}={\frac{2}{\pi^2}}
{\int_{0}^{\pi/2}}{\int_{0}^{\pi}}{\int_{0}^{\pi/2}}\pi r_V^2 \, Q_{\rm pol}\,
f(\xi, \beta) \, \cos 2{\psi} \, d{\varphi} d{\omega} d{\beta} \,.
\label{eq4}
\end{equation}
Here, $r_V$ is the radius of a sphere with the same volume
as spheroidal grain,
${\beta}$ is the precession-cone angle for
the angular momentum {\bf J} which spins around the magnetic field
direction {\bf B}, ${\varphi}$ the spin angle, ${\omega}$  the precession
angle (see Fig.~1 in \cite{nvv1}).
The quantities $Q_{\rm ext}=(Q_{\rm ext}^{\rm TM} + Q_{\rm ext}^{\rm TE})/2$
and  $Q_{\rm pol}=(Q_{\rm ext}^{\rm TM} - Q_{\rm ext}^{\rm TE})/2$ are,
respectively, the extinction and polarization efficiency factors for
the non-polarized incident radiation,
$f(\xi,r_V)$ is the cone-angle distribution.

We consider so-called imperfect Davis--Greenstein
(IDG) orientation  \cite{dg51}.
The IDG mechanism is described by the function
${f}(\xi, \beta)$ which depends on the alignment parameter $\xi$ and
the angle $\beta$
\be
{f}(\xi, \beta) = \frac{\xi \sin \beta}{(\xi^2 \cos^2 \beta  + \sin^2 \beta)^{3/2}}.
\label{idg} \ee
The parameter $\xi$ depends on the
particle size $r_V$, the imaginary part of the grain
magnetic susceptibility
$\chi''$ ($=\varkappa \omega_{\rm d} /T_{\rm d}$, where $\omega_{\rm d}$ is the angular velocity
of grain), gas  density $n_{\rm g}$, the strength of magnetic field $B$
and dust ($T_{\rm d}$) and gas ($T_{\rm g}$) temperatures\index{Dust grains!temperature}
\be
\xi^2  = \frac{r_V +\delta_0 (T_{\rm d}/T_{\rm g})}{r_V +\delta_0},
\,\,\,\,\,\,
{\rm where}
\,\,\,\,\,\,
\delta_0^{\rm IDG} = 8.23\,10^{23} \frac{\varkappa B^2}{n_{\rm g} T_{\rm g}^{1/2} T_{\rm d}}\,\mkm.
\label{xi} \ee
The angle $\psi$ in Eq.~(\ref{eq4}) is expressed
via the angles $\varphi, \omega, \beta$ and $\Omega$
(angle between the line of sight and the magnetic field,
$0\degr \leq \Omega \leq 90\degr$).
Note that for the case of the perfect DG orientation (PDG, perfect rotational
or 2D  orientation)
the major axis of a non-spherical particle always lies in the same plane.
For PDG,  integration in Eqs.~(\ref{eq3}), (\ref{eq4})
is performed over the spin
angle $\varphi$ only.

As usual, the particles of different sizes are considered.
In this case the cross-sections
$ \langle C_{\rm ext} \rangle_{\lambda}$ and
$ \langle C_{\rm pol} \rangle_{\lambda}$ are obtained after the
averaging over size distribution function $n(r_V)$
$$
{\int_{r_{V, \min}}^{r_{V, \min}}} \langle C \rangle_{\lambda}(r_V)
\,n(r_V) \, d{r_V} \,,
$$
where $r_{V, \min}$ and $r_{V, \max}$ are the lower and upper
cutoffs, respectively.

We choose the power-law size distribution function
$$
n(r_V) \propto r_V^{-q}\,,
$$
which was often used for the modelling of interstellar
extinction \cite{mrn}, \cite{dl}.
The average Galactic extinction curve can be reproduced
using the mixture of carbonaceous (graphite) and silicate particles
(in almost equal proportions)
with parameters: $q=3.5$, $r_{V, \min}\approx 0.001 \,\mu$m
and $r_{V, \max} \approx 0.25 \,\mu$m.

\section{Results and discussion}

Our calculations have been performed for prolate and oblate
rotating spheroids of several sizes  and aspect ratios.
The particles consisting of astronomical silicate (astrosil)
and amorphous carbon (AC1) were considered.
The optical properties of spheroids were derived
using the solution to the light scattering problem by the
separation of variables method \cite{vf93} and modern
treatment of spheroidal wave functions \cite{vf04}.

Below we briefly discuss extinction
by spheroids and  in sufficient detail polarizing efficiency.
The comparison with observations of two stars
is performed in Sect.~\ref{comp}.

\subsection{Extinction curve}

Extinction (and sometimes polarization) produced by spheroids
was studied by Rogers and Martin \cite{rm79},
Onaka \cite{on80}, Vaidya et al. \cite{va84}, Voshchinnikov \cite{v90}
and more recently by Gupta et al. \cite{va05}.
Mostly the particles with the aspect ratio $a/b\leq 2$ were considered.

\begin{figure}[htb]
\centerline{
\resizebox{\hsize}{!}{\includegraphics{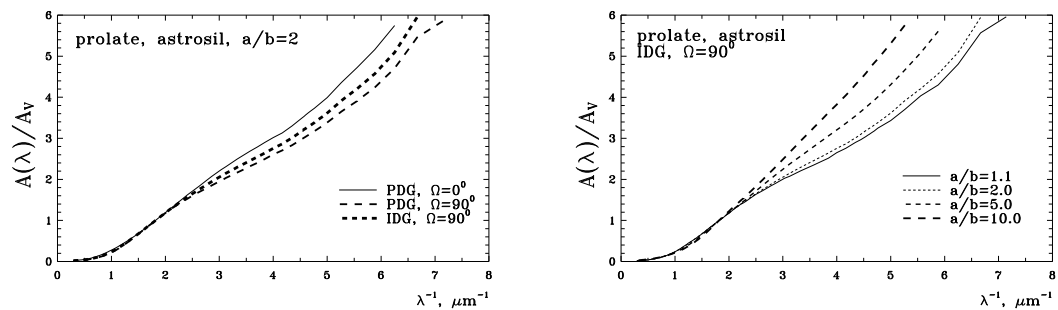}}
}
\caption
{Normalized extinction for ensembles of aligned
prolate spheroids from astrosil. Spheroids have
power-law size distribution with parameters:
$r_{V, \min}=0.001 \,\mu$m, $r_{V, \max} = 0.25 \,\mu$m and
$q=3.5$; $\delta_0=0.196\,\mu$m.
The effect of variations of
particle degree and direction of alignment (left panel)
and particle shape (right panel) is illustrated.
}
\label{ext}
\end{figure}

The parameters of our model (see Sect.\ref{mod}) are:
the type (prolate or oblate) and  composition of spheroids,
the size $r_V$ or the parameters  of the size distribution
$r_{V, \min}$, $r_{V, \max}$, $q$, the degree
specified through parameter $\delta_0$ and direction of
alignment $\Omega$.
Figure~\ref{ext} illustrates how  variations of
particle shape and alignment influence on the
normalized extinction $A(\lambda)/A_V$.
The value of $\delta_0$ for IDG orientation is typical of
diffuse interstellar medium  \cite{nvv1}.
It is seen that the difference in  curves
are in evidence only in the UV. Prolate spheroids
produce larger extinction when the angle of alignment
reduces from $\Omega=90\degr$
(magnetic field is perpendicular to the line of site)
to $\Omega=0\degr$
(magnetic field is parallel to the line of site;
Fig.~\ref{ext}, left panel).
For oblate particles the dependence on $\Omega$ is
opposite. However, only for PDG these changes are at
a detectable rate. Extinction also grows when the particles
(prolate and oblate) become more
elongated or flattened (see Fig.~\ref{ext}, right panel).

Variations of parameters of size distribution
($r_{V, \min}$, $r_{V, \max}$, $q$) on extinction
are similar to those for spheres are not discussed here
(see, e.g., \cite{vi93}).

\subsection{Polarizing efficiency: theory}

Theoretical dependence of the polarizing efficiency of the interstellar
medium was discussed several times for particles of constant
refractive index (see, e.g., \cite{rm79}, \cite{on80}).
The wavelength dependence of $P/A$ was considered
in \cite{lig97} for infinite and finite cylinders and
in \cite{nvv2} for perfectly aligned spheroids of single size.
Here, we analyse firstly the dependence $P(\lambda)/A(\lambda)$ for
single size particles  and then for ensembles with a size distribution.

\subsubsection{Single size}

\begin{figure}[htb]
\centerline{
\resizebox{\hsize}{!}{\includegraphics{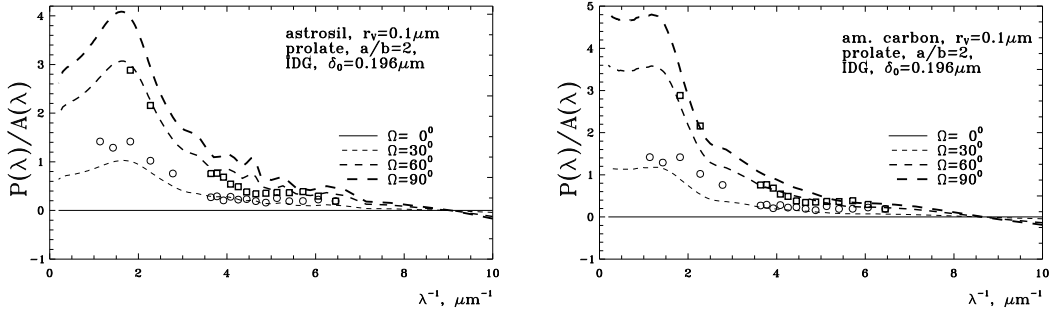}}
}
\caption
{Wavelength dependence of polarizing   efficiency  for
homogeneous rotating spheroidal particles of astronomical silicate and
amorphous carbon.
The effect of variations of particle composition and direction
of alignment is illustrated.
The open circles and squares show the observational data
for stars  HD~24263 and HD~99264, respectively.
}
\label{fig3}
\end{figure}

\begin{figure}[htb]
\centerline{
\resizebox{\hsize}{!}{\includegraphics{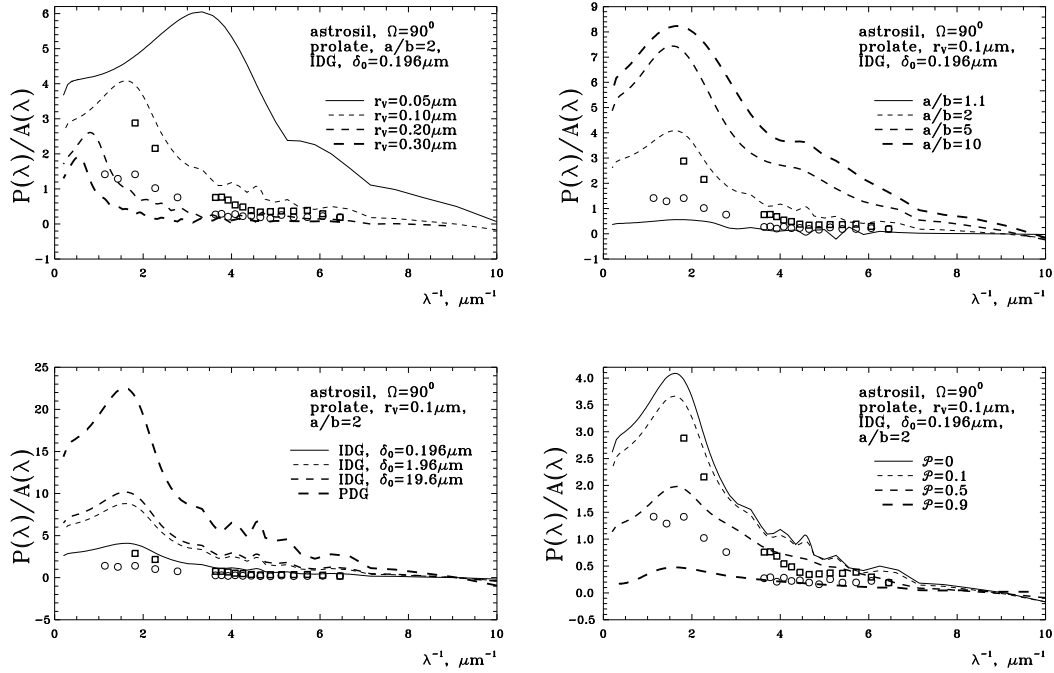}}
}
\caption
{Wavelength dependence of polarizing   efficiency  for
homogeneous rotating spheroidal particles of astronomical silicate.
The effect of variations of particle size, shape, porosity and degree
of alignment is illustrated.
The open circles and squares show the observational data
for stars  HD~24263 and HD~99264, respectively.
}
\label{fig4}
\end{figure}

Some results for prolate particles
are plotted in Figs.~\ref{fig3} and ~\ref{fig4}.
They show the polarizing   efficiency
in the wavelength range from the infrared to far ultraviolet.
The  dependence $P(\lambda)/A(\lambda)$  observed
for two stars is given for comparison.
We made calculations for compact grains and
for porous grains  using the Bruggeman mixing rule
for refractive index and particles of same mass as compact ones.
Note that calculated efficiencies can be considered as upper
limits because some populations of grains (spherical, non-oriented)
may  contribute into extinction but not to polarization.


From Figs.~\ref{fig3}, ~\ref{fig4} one can conclude that
the wavelength dependence of polarizing efficiency is  mainly
determined by the particle composition and size.
Variations of other parameters influence on the value of efficiency
(the dependence of $P/A$ is scaled). A growth of the
efficiencies $P(\lambda)/A(\lambda)$ takes the place if we increase
angle $\Omega$ (direction of alignment deviates widely from the line
of site),
parameter $\delta_0$ (degree of alignment) or
aspect ratio $a/b$ (consider more elongated or flattened particles)
and decrease particle porosity (volume fraction of vacuum)
${\mathcal P}$ or particle size $r_V$.
It is also evident that the contribution of particles
with size $r_V \approx 0.1 \,\mu$m to observed polarization must
be rather important and the polarization produced by perfectly aligned
particles significantly exceeds the observed one.

\subsubsection{Size distribution}

\begin{figure}[htb]
\centerline{
\resizebox{\hsize}{!}{\includegraphics{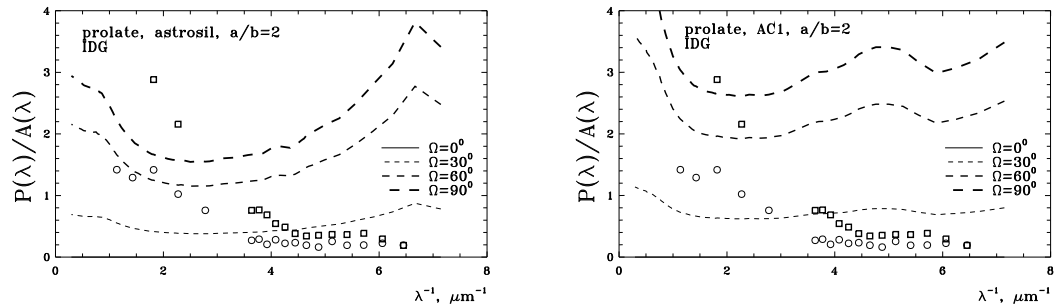}}
}
\caption
{Polarizing  efficiency
ensembles of aligned prolate spheroids with $a/b=2$
from astrosil (left panel) and amorphous carbon (right panel).
Spheroids have
power-law size distribution with parameters:
$r_{V, \min}=0.001 \,\mu$m, $r_{V, \max} = 0.25 \,\mu$m and
$q=3.5$; $\delta_0=0.196\,\mu$m.
The effect of variations of
particle composition and direction of alignment is illustrated.
The open circles and squares show the observational data
for stars  HD~24263 and HD~99264, respectively.
}
\label{pt2}
\end{figure}

\begin{figure}[htb]
\centerline{
\resizebox{\hsize}{!}{\includegraphics{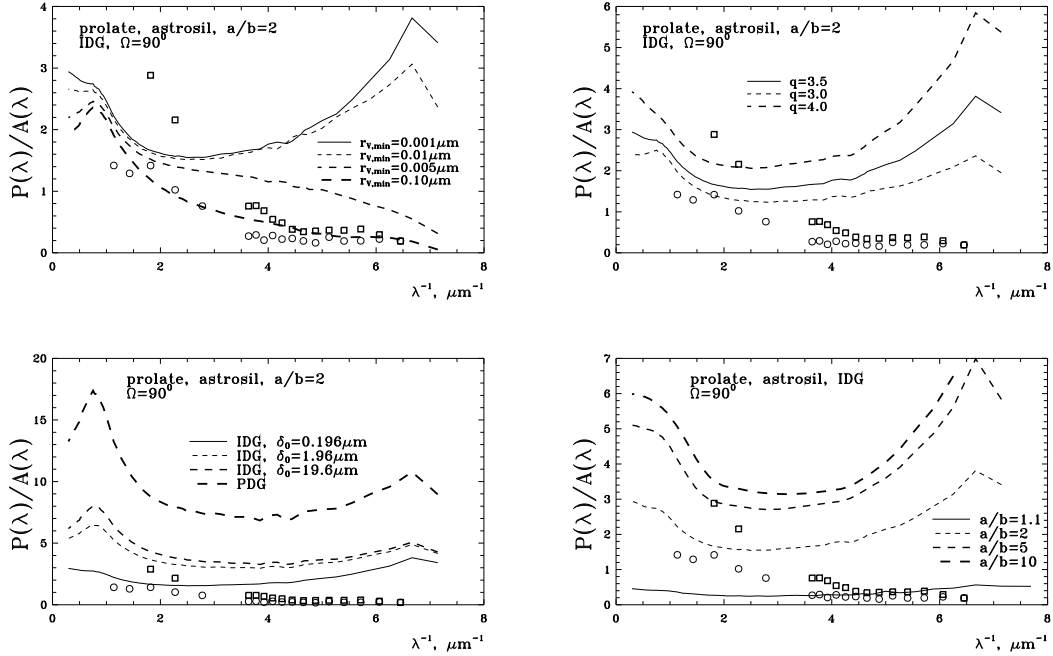}}
}
\caption
{Polarizing  efficiency
ensembles of aligned prolate spheroids from astrosil.
Spheroids have power-law size distribution, $\Omega=90\degr$.
The effect of variations of parameters of size distribution,
degree of alignment and particle shape is illustrated.
The open circles and squares show the observational data
for stars  HD~24263 and HD~99264, respectively.
}
\label{ppp}
\end{figure}

We also calculated the wavelength dependencies of $P/A$
separately for spheroids from AC1 and astrosil with standard
power-law size distribution
($r_{V, \min}=0.001 \,\mu$m, $r_{V, \max} = 0.25 \,\mu$m and
$q=3.5$) and considered the effect of variations different
model parameters. The results are shown in
Figs.~\ref{pt2}, and~\ref{ppp}.
Note again that  the value of $P/A$ depends on the degree and
direction of alignment and the shape of particles while
the particle composition and size strongly influences on the
wavelength dependence of polarizing efficiency.


As follows from these Figures,
theoretical polarizing efficiency more or less resembles
the observational dependencies only if the contribution
of small particles into polarization is depressed
($r_{V, \min} > 0.05 \,\mu$m or $q <3$, see Fig.~\ref{ppp}).
It is well known for a long time that the polarization of
forward transmitted radiation is determined by particles of
small sizes while large particles produce no polarization independent
of their shape (see discussion in \cite{gre68} and  \cite{vihf00}).

The influence of small particles on $P/A$ is reduced if:
a) the particles are absent in the interstellar cloud,
b) the small particles are less oriented or
c) the shape of small particles is closer to spherical.
The reason a) must manifests itself by the depressed UV
extinction while the reasons b) and c) would be
difficult to distinguish.

The  relationships found can be used for more
detailed comparison of the theory with observations.

\subsection{Comparison with observations}\label{comp}

We calculated the normalized extinction $A(\lambda)/A_V$
and the polarizing  efficiency and compared the results with
observations of stars mentioned in Sect.~\ref{obs}.
In order to fit the UV bump on the extinction curve a small
amount of graphite spheres with radius $r_V \approx 0.015\,\mu$m
was added to the mixture of silicate and carbon spheroids.

According to the model presented in Sect.~\ref{mod},
we adopted that the same particles produce extinction
and polarization and the alignment of silicate and carbon particles
is identical. At the same time, parameters of size distribution
for silicate and carbon can differ.

\begin{figure}[htb]
\centerline{
\resizebox{\hsize}{!}{\includegraphics{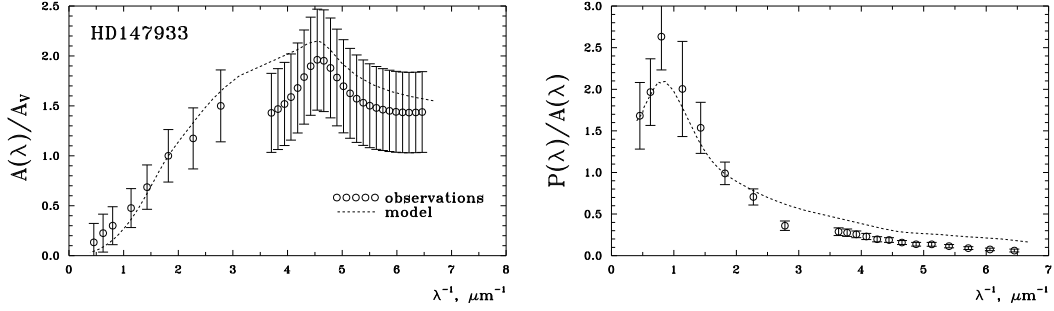}}
}
\caption
{Normalized extinction (left panel)
and the polarizing  efficiency (right panel)
in the direction of the star HD~147933.
The error bars show 1$\sigma$ uncertainty.
Dashed curves show the results of calculations for model with
prolate spheroids with $a/b=5$ from amorphous carbon (24\%) and
astrosil (73\%) and graphite spheres (3\%). Spheroids have
a power-law size distribution with the parameters:
amorphous carbon,
$r_{V, \min}=0.03 \,\mu$m, $r_{V, \max} = 0.15 \,\mu$m, $q=1.5$;
astrosil,
$r_{V, \min}=0.08 \,\mu$m, $r_{V, \max} = 0.25 \,\mu$m, $q=1.5$.
Alignment parameters are: $\delta_0=0.5\,\mu$m, $\Omega=35\degr$.
}
\label{fig147}
\end{figure}

\begin{figure}[htb]
\centerline{
\resizebox{\hsize}{!}{\includegraphics{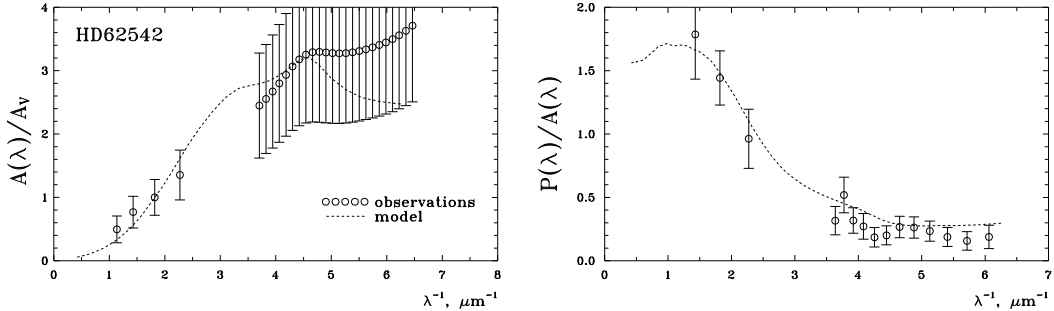}}
}
\caption
{Normalized extinction (left panel)
and the polarizing  efficiency (right panel)
in the direction of the  star HD~62542.
The error bars show 1$\sigma$ uncertainty.
Dashed curves show the results of calculations for model with
prolate spheroids with $a/b=2$ from amorphous carbon (85\%) and
astrosil (10\%) and graphite spheres (5\%). Spheroids have
a power-law size distribution with the parameters:
amorphous carbon,
$r_{V, \min}=0.07 \,\mu$m, $r_{V, \max} = 0.10 \,\mu$m, $q=2.7$;
astrosil,
$r_{V, \min}=0.005\,\mu$m, $r_{V, \max} = 0.25 \,\mu$m, $q=2.7$.
Alignment parameters are: $\delta_0=0.5\,\mu$m, $\Omega=40\degr$.
}
\label{fig625}
\end{figure}

The results of comparison are shown for two stars
in Figs.~\ref{fig147} and~\ref{fig625}.

The coincidence of the theory with observations is rather
good. The better agreement can be achieved if we assume that
the alignment parameters for silicate and carbonaceous grains
differ. For example, it is possible to consider non-aligned
carbonaceous grains (as it was made in \cite{m79}).
Evidently, more attention should be given to modern alignment
mechanisms.

These improvements of the model require to invoke additional
observational data such as dust-phase abundances in the directions
of considered stars, information about magnetic fields, etc.
Otherwise, the modelling cannot give non-ambiguous results.
In particular, the important feature is the wavelength where
the ratio $P/A$ reaches a maximum (see Figs.~\ref{fig3}--\ref{ppp}).
In order to find this quantity we need the observations of
polarization in the near-IR (J, H, K bands) which are absent for
some considered stars (Fig.~\ref{fig2}).

\section{Conclusions}


In the frame of the model of partially aligned spheroidal
grains we studied the wavelength dependence of the
       ratio of the linear polarization degree to extinction
       (polarizing efficiency) $P(\lambda)/A(\lambda)$.

The main results  can be summarized as follows.

1. It is found that the wavelength dependence of $P(\lambda)/A(\lambda)$
is mainly determined by the particle composition and size.

2. The values of $P(\lambda)/A(\lambda)$ depend on the
particle shape, degree and direction of alignment.

3. The  modelling of the only wavelength dependence of
polarizing efficiency does not allow one to determine
all parameters dust ensemble. Therefore,
interpretation of the observations must include consideration
of extinction.

\noindent{\bf Acknowledgements}

We are thankful to Vladimir Il'in for careful reading of manuscript.
The work was partly supported by
grants NSh 8542.2006.2, RNP 2.1.1.2152 and RFBR 07-02-00831
of the Russian Federation.

\end{document}